\documentclass[5p]{elsarticle}
\pdfoutput=1
\usepackage[utf8]{inputenc}
\usepackage[T1]{fontenc}
\usepackage{hyperref}
\usepackage{color}
\usepackage{graphicx}
\usepackage{units}
\usepackage{amsmath}
\usepackage{float}
\usepackage{epstopdf}
\usepackage{subcaption}
\usepackage{amssymb}
\usepackage{pifont}
\usepackage{textcomp}
\usepackage[usenames,dvipsnames]{xcolor}
\usepackage{ulem}
\usepackage{etoolbox}
\usepackage{booktabs}
\usepackage{multirow}
\usepackage{amsmath}

\newcommand{\angstrom}{\text{\normalfont\AA}}
\newcommand{\etal}{\textit{et al}. }

\graphicspath{{pictures/}}

\hypersetup{colorlinks = true, citecolor = blue, breaklinks = true}

\begin{document}

\title{Water Oxidation Chemistry of Oxynitrides and Oxides:\\Comparing NaTaO$_3$ and SrTaO$_2$N}
\date{\today}
\author[dcb]{Hassan Ouhbi}
\author[dcb]{Ulrich Aschauer\corref{cor}}
\ead{ulrich.aschauer@dcb.unibe.ch}
\address[dcb]{Department of Chemistry and Biochemistry, University of Bern, Freiestrasse 3, CH-3012 Bern, Switzerland}
\cortext[cor]{Corresponding author}

\begin{abstract}
The oxygen evolution reaction (OER) plays an important role in evaluating a photocatalyst and to understand its surface chemistry. In this work we present a comparative study of the OER on the oxide NaTaO$_3$ (113) surface and the oxynitride SrTaO$_2$N (001) surface. Oxynitrides are highly promising photocatalysts due to their smaller band gap and resulting better visible light absorption compared to oxides but our knowledge about their surface structure and chemistry is still very limited. With the goal to compare the surface chemistry of oxides and oxynitrides, we perform density functional theory calculations to obtain the free energy changes associated with the OER reaction steps. For the OER at the Ta site of the clean surfaces, our results predict the rate-limiting step for both materials to be the formation of the *OOH intermediate, with a larger overpotential for the oxide than the oxynitride (1.30 V vs 1.01 V). The Na site is found to be more active than the Ta site on the oxide surface with an OER overpotential of 0.88 V, whereas the OER at the Sr site on the oxynitride has an overpotential of 1.14 V. For the A sites, contrary to the Ta site, the deprotonation of *OH was found to be the rate-limiting step. Computed Pourbaix diagrams show that at relevant (photo)electrochemical conditions all surfaces are covered with oxygen adsorbates. Oxygen adsorbates at A (Na, Sr) sites are however found to couple and desorb as O$_2$, leaving these sites empty under typical operating conditions. Following this desorption, we find the OER to proceed by the conventional *OOH mechanism on the SrO termination of the oxynitride but by a direct coupling of neighbouring *O at Na sites on the oxide surface. This coupling mechanism on the oxide has the smallest overpotential of 0.79 V compared to 0.88 V for the oxynitride, implying that the oxide is a better OER catalyst. Since it however absorbs light only in the UV part of the solar spectrum this leads to a tradeoff between light absorption and the catalytic activity.\\\\
Keywords: Oxynitride; Oxide; Surface chemistry; Oxygen evolution; Density functional theory
\end{abstract}

\maketitle

\section{Introduction}

\begin{figure}[t]
\centering
 \includegraphics{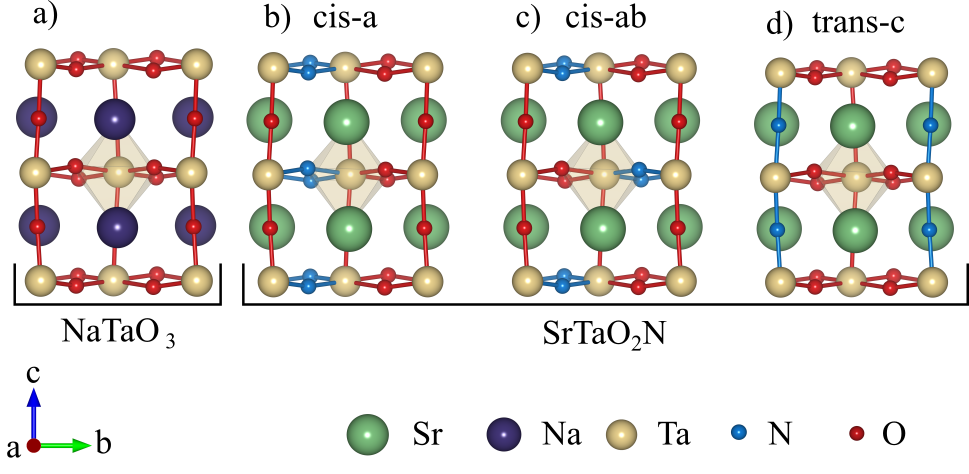}
\caption{Bulk structures of a) the oxide NaTaO$_3$ and b)-d) the oxynitride SrTaO$_2$N with b) \textit{cis-a}, c) \textit{cis-ab} and d) \textit{trans-c} anion orders.}
\label{fig:bulk_structures}
\end{figure}

Photocatalytic water splitting is considered to be a promising route for the production of clean hydrogen fuel. This reaction was first observed for TiO$_2$ under ultraviolet (UV) light by Fujishima and Honda \cite{Fujishima}. Since then various materials have been examined as potential photocatalysts \cite{Osterloh2008}. An ideal photocatalyst must be able to absorb light not only in the UV but also in the visible part of the solar spectrum, which limits the band gap to below $\sim$ 3 eV. In order to enable overall water splitting, a photocatalyst must further have valence-band and conduction-band edges that straddle the redox potential for H$^+$/H$_2$ and H$_2$O/O$_2$ to enable the hydrogen evolution reaction (HER) and the oxygen evolution reaction (OER) respectively. Perovskite structured tantalates with the general formula ATaO$_3$, such as NaTaO$_3$ (see Fig. \ref{fig:bulk_structures}a) or KTaO$_3$, have received much attention for this application due to their high quantum efficiency \cite{Grabowska2016}. Unfortunately their wide band gap (NaTaO$_3$ = 4.1 eV, KTaO$_3$ = 3.6 eV) limits their light absorption to UV radiation \cite{Grabowska2016,Modak2016}.

The chemical and structural flexibility of the perovskite structure allows exploiting different strategies to narrow the band gap, one possibility being substitution on the cation or anion sites. Perovskite tantalate oxynitrides (Fig. \ref{fig:bulk_structures}b-d) with the general formula ATaO$_2$N are members of the latter class of materials, where a partial substitution of oxygen by nitrogen leads to a smaller band gap (SrTaO$_2$N = 2.1 eV) due to the lower electronegativity of nitrogen compared to oxygen and the resulting upwards shift of the valence-band edge \cite{Takata2016}. While the bulk structure and electronic properties of perovskite oxides and oxynitrides have been explored in computational screening investigations \cite{Castelli2012,Wu2013,Pena2001}, only few studies have been devoted to understand their surface chemistry \cite{Man2011,Montoya2015a,Montoya2018}.

\begin{figure}
\centering
 \includegraphics{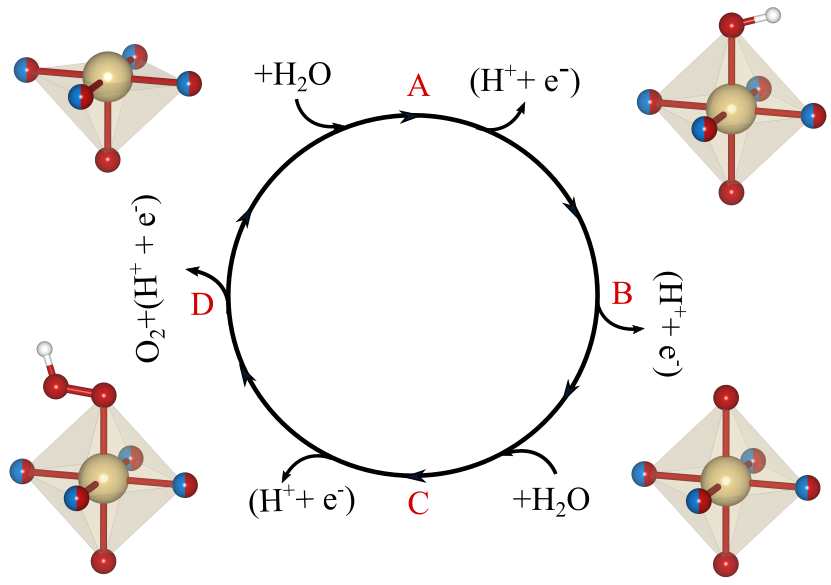}
\caption{Primary oxygen evolution-reaction mechanism considered in this work.}
\label{fig:OER}
\end{figure}

The OER is considered the bottleneck in water splitting and therefore most appropriate to evaluate and compare the activity of different catalyst surfaces \cite{Whitesides2007,Hong2015}. Various mechanism have been proposed for the OER and it has been demonstrated that the most favourable pathway is material and environment dependent \cite{Zhang2016a}. In this study we initially adopt the mechanism schematically shown in Fig. \ref{fig:OER}, which is a succession of four proton-coupled electron transfer (PCET) steps \cite{Norskov2004}. First, a water molecule is deprotonated to form an adsorbed hydroxyl (step A), which then loses a second proton to become an adsorbed oxygen (step B). A further deprotonation involving a second water molecule leads to the formation of a OOH hydroperoxyl group (step C), which, after a final deprotonation, is released as O$_2$ (step D).

N\o rskov \etal paved the way to computationally assess the thermodynamics of this mechanism by developing a scheme based on density functional theory (DFT) calculations \cite{Norskov2004}. A key component in this approach is the computational standard hydrogen electrode (SHE)\cite{Norskov2004}, which assumes that the chemical potential of H$^{+}$ $+$ e$^{-}$ is equivalent to that of $\frac{1}{2}$H$_{2}$ at standard conditions (pH=0, p=1 bar, and T=298 K). Hence, at standard conditions and in the absence of an applied bias potential (U$_b$=0 V) the reaction HA$^{*}$ $\longrightarrow$ A $+$ H$^{+}$ $+$ e$^{-}$ is equivalent to HA$^{*}$ $\longrightarrow$ A $+$ $\frac{1}{2}$H$_{2}$. The application of an external potential bias changes the chemical potential of the electrons by $-eU_b$, while the free energy of the proton reservoir is changed by $-k_B ln(10)\cdot \mathrm{pH}$ at a given pH. Consequently, the free energy change $\Delta$G for each PCET step is calculated as
\begin{equation}
\begin{split}
\Delta G(U_b,pH) & = \Delta E + (\Delta ZPE - T\Delta S) \\ 
& - eU_b - k_BT\cdot \ln(10)\cdot \mathrm{pH} 
\end{split}
\label{eq:deltaG}
\end{equation}
where $\Delta$E is the DFT-calculated reaction energy and $\Delta$ZPE and $\Delta$S are the changes in zero point energy and entropy of the reaction intermediates respectively. The thermodynamic overpotential $\eta$, which we use to characterise the OER activity of a given surface, is defined as the potential for which all PCET steps have $\Delta G(U_b,pH)$ smaller than zero relative to the equilibrium potential of 1.23 V:
\begin{equation}
	\eta = \max(\Delta G)/e - 1.23 \mathrm{V}
\end{equation}

This method has been adopted in different theoretical studies to investigate a variety of materials \cite{Rossmeisl2007, Nguyen2015}. It was found that the difference in adsorption free energies $\Delta$G$_O$-$\Delta$G$_{OH}$ can often be used as an universal descriptor for the reactivity of a given surface. This descriptor shows a good correlation with the overpotential, giving rise to Sabatier volcano plots \cite{Trasatti1972} and provides a theoretical framework to compare different materials \cite{Man2011}. Recently, Montoya \etal have studied the OER on (100) surfaces of different oxide and oxynitride perovskites. According to their results, SnTiO$_3$ needs the largest overpotential and has the strongest oxygen binding energy, while CaTaO$_2$N is the most active \cite{Montoya2015a}. Furthermore, it is important to identify the dominant OER mechanism under (photo)electrochemical conditions as the surface adsorbate coverage changes with pH and the applied potential. DFT calculations have been used to construct surface Pourbaix diagrams, based on which the most relevant reaction mechanism can be identified, leading to more accurate results of the activity compared to experiment \cite{Russell2008}.

Herein, we present a comparative study of the perovskite oxide NaTaO$_3$ and the oxynitride SrTaO$_2$N in terms of their surface water-oxidation chemistry. There is still a debate whether the valence-band edge of SrTaO$_2$N lies above the O$_2$/H$_2$O level \cite{Wu2013}, which would experimentally lead only to H$_2$ evolution but no O$_2$ evolution \cite{Yamasita2004} or below that level \cite{Balaz2013}. Notwithstanding this fact, our goal here is to study the effect of partial nitrogen substitution on the surface structure and chemistry of chemically similar oxides and oxynitrides. We compute the free-energy profile of the OER on the (001) and (113) surface respectively of these two materials and determine the overpotential at the A site (Na, Sr) and the B site (Ta). We find the B site on the oxynitride (001) surface to be more active than its counterpart on the oxide (113) surface. The A site is predicted to be the most active for the oxide but less active for the oxynitride.  An alternative OER mechanism involving coupling of oxygen adsorbates is found to be more favourable on the oxide surface under electrochemical conditions.

\section{Methods}

All reactions energies ($\Delta E$ in equation \ref{eq:deltaG}) are calculated using density functional theory, as implemented in the Quantum ESPRESSO package \cite{Giannozzi2009}, using the Perdew-Burke-Ernzerhof (PBE) exchange correlation functional \cite{Perdew1996}. Inclusion of a Hubbard U correction \cite{Anisimov1991} on the Ta 5d states was found to have no significant effect on the electronic structure and lattice parameters (see supporting information Tab. S1) and was therefore omitted. Ultrasoft pseudopotentials \cite{Vanderbilt1990} with Na(2p$^6$ 3s$^1$), Sr(4p$^6$5s$^2$), Ta(6s$^2$5d$^3$), O(2s$^2$2p$^4$) and N(2s$^2$2p$^3$) valence states were used to describe electron-nuclear interactions and wave functions were expanded in plane waves up to a kinetic energy cutoff of 40 Ry combined with 320 Ry for the augmented density.   

We started from the experimental orthorhombic NaTaO$_3$ \textit{Pbnm} perovskite structure \cite{Ennedy2006} and the SrTaO$_2$N \textit{I4/mcm} perovskite structure \cite{Clarke2002}, which were fully relaxed at the PBE level of theory. A 8 Na$_4$Ta$_4$O$_{12}$ layer thick (113) slab (lateral dimensions 11.044 $\times$ 19.133 \AA) was constructed for the oxide, while a (001) slab (lateral dimensions 8.182 $\times$ 8.182 \AA) with 8 TaON/SrO layers is used for the oxynitride. The rationale behind the selection of these specific surface orientations will be discussed below. All slabs are separated by a vacuum of 15 $\angstrom$ along the surface normal direction and the bottom two atomic layers were fixed at bulk positions. A k-point mesh with dimension 2$\times$1$\times$1 and 4$\times$4$\times$1 is used for the NaTaO$_3$ and SrTaO$_2$N surfaces respectively \cite{Pack1977}. A dipole correction \cite{Bengtsson1999} was introduced along the z-direction for all slab calculations. A threshold of 0.001 eV$\angstrom^{-1}$ and 10$^{-6}$ eV for forces and total energies respectively is used during structural relaxations.

Vibrational normal modes and associated frequencies ($\nu_i$) of the adsorbates were computed using the frozen-phonon method as implemented in the PHONOPY code \cite{Togo2015a} using atomic displacements of 0.01 $\angstrom$. The zero-point energy is obtained by summing the contributions of the $N_\mathrm{modes}$ normal modes of the adsorbate:
\begin{equation} 
ZPE = \sum_{i=1}^{N_\mathrm{modes}} \frac{1}{2}h\nu_{i}, 
\end{equation} 
where $h$ is Planck's constant. This approach was found to give equivalent results for zero-point energy differences ($\Delta$ZPE in equation \ref{eq:deltaG}) compared to a full treatment involving also vibrations of slab atoms, implying that these slab contributions do not change significantly and cancel. For entropy contributions ($\Delta$S in equation \ref{eq:deltaG}) adsorbed molecules were assumed to have zero entropy, while those for gas-phase H$_2$O and H$_2$ molecules were taken from the JANAF tables at the gas-liquid equilibrium (P=0.035 bar and T=298.15 K) \cite{Chase}.
 
\section{Results and Discussion}

\subsection{Bulk structure and anion order}

We first investigated the O/N anion order in the bulk oxynitride structure, comparing different \textit{cis} and \textit{trans} arrangements of the N ions. We find that the \textit{cis-a} order (Fig. \ref{fig:bulk_structures}b) is the energetically most favourable bulk anion order, the \textit{trans-c} (Fig. \ref{fig:bulk_structures}c) and \textit{cis-ab} (Fig. \ref{fig:bulk_structures}d) anion orders being 0.27 eV and 0.03 eV per formula unit respectively higher in energy. This is in agreement with previous experimental findings, which found the \textit{cis} anion order to be stable up to temperatures of 2000 $^\circ$C \cite{Clark2013}. The \textit{cis-a} anion order lowers the symmetry from the room temperature \textit{I4/mcm} structure with a$^0$a$^0$c$^+$ octahedral rotations \cite{Glazer:1972eb} to a structure with space group \textit{Pm} and a$^-$a$^-$c$^+$ octahedral rotations. Indeed, this symmetry reduction was reported by both experimental and theoretical studies and was suggested to result from correlation with rotations of the TaO$_4$N$_2$ octahedra \cite{Yang2011b}. For the oxide on the other hand the structure keeps the same a$^-$a$^-$c$^+$ octahedral rotation pattern as in the experimental structure.

The lattice parameters of the relaxed unit cells are comparable to the experimental ones with less than 1\% and 2\% deviations for the oxide and the oxynitride respectively (see supporting information Tab. S1). The  partial density of states (PDOS) shown in supporting information Fig. S1 reveals that the bottom of the conduction band is mainly formed by Ta 5\textit{d} orbitals for both the oxide and the oxynitride. The top of the valence band in perovskite oxides consists primarily of O 2\textit{p} states while for the oxynitride it is composed of N 2\textit{p} states, which leads to an upwards shift of the valence-band edge compared the oxide. Consequently we predict a smaller band gap for the oxynitride compared to the oxide (see supporting information Tab. S1), highlighting the enhanced visible-light absorption of the former class of materials. We note, however, that as expected for semi-local DFT calculations, the absolute value of the band gap is underestimated compared to experiment.

\subsection{Surface structure}

\begin{figure}
\centering
 \includegraphics{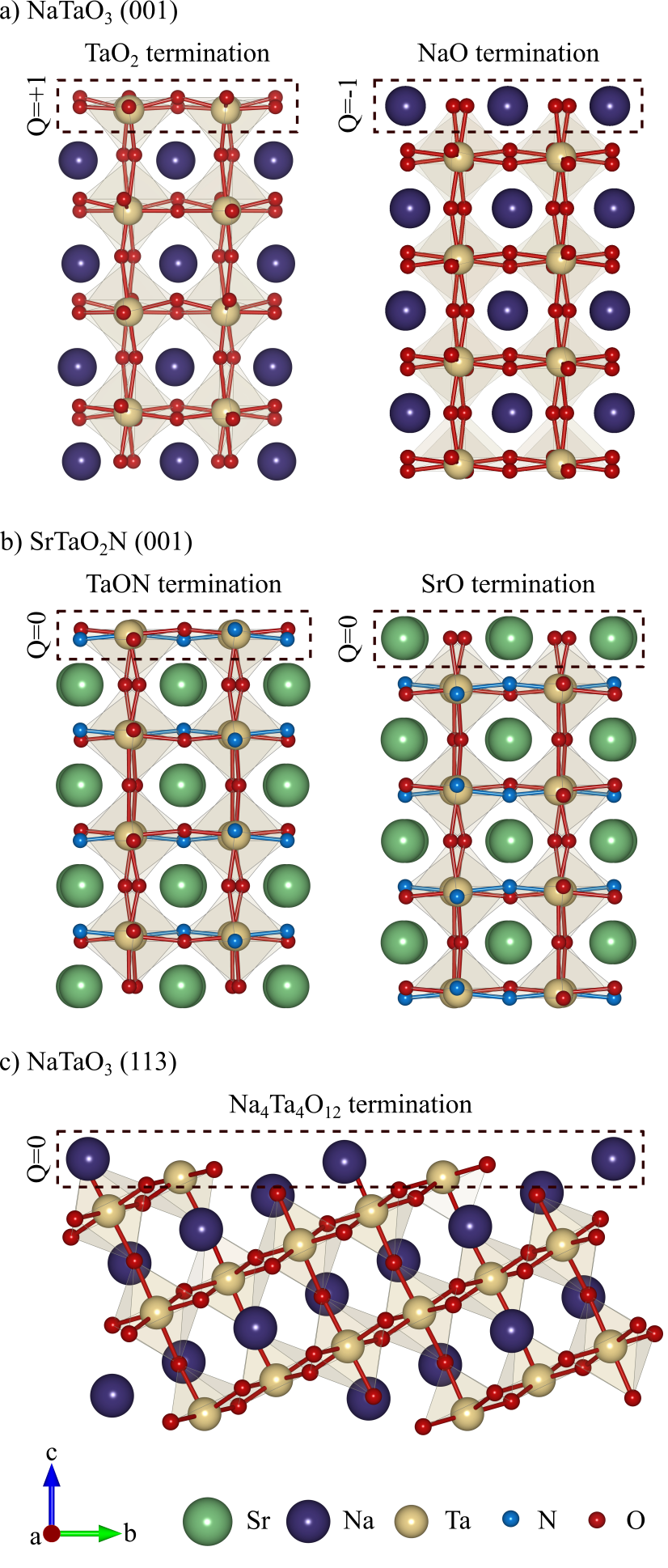}
\caption{Slab models and charge per layer Q of a) NaTaO$_3$ (001), b) SrTaO$_2$N (001) and c) NaTaO$_3$ (113).}
\label{fig:slabs}
\end{figure}

Cleaving NaTaO$_3$ perpendicular to the [001] direction, results in a (001) surface with alternating NaO and TaO layers with formal charges of -1e and +1e respectively as shown in (Fig. \ref{fig:slabs}a). Such a surface is polar and unstable due to its diverging electrostatic energy \cite{Noguera2000}. Such surfaces will stabilise by introducing compensating charges either by electronic reconstruction, adsorption of charged species or ionic reconstruction \cite{Noguera2013,Deacon-Smith2014a}. Here, we follow an alternative strategy and investigate the (113) surface, which is formed by stacking of charge-neutral Na$_4$Ta$_4$O$_{12}$ layers and hence non-polar (Fig. \ref{fig:slabs}c). Unlike for NaTaO$_3$, the SrTaO$_2$N (001) surface cleaved from the most stable \textit{cis-a} anion-ordered bulk is non-polar since it consists of alternating charge-neutral TaON and SrO layers as shown in (Fig. \ref{fig:slabs}b).

From the layer-resolved electronic density of states shown in supporting information Fig. S2, we observe that, in agreement with the bulk density of states, the conduction-band for the surfaces of both materials is mainly composed of Ta 5\textit{d} states, while the valence-band is dominated by O 2\textit{p} states in the case of oxide, and N 2\textit{p} states for the oxynitride. In the topmost layer of the TaON-terminated oxynitride surface, we observe a shift of the N states to higher energies compared to subsurface layers and bulk states, while for the oxide surfaces, the O 2\textit{p} states, which mainly form the valance band are shifted to slightly lower energy compared to the bulk states. The highest occupied oxygen states in SrO layers do not exhibit such a shift independent of the surface termination. Sr and Na states have negligible contribution, except for the topmost layer of SrO-terminated surface where some contribution is observed close to the Fermi level.   

\subsection{Water oxidation chemistry}

The oxygen evolution reaction is studied on both terminations (TaON and SrO) of the oxynitride (001) surface, and on the oxide (113) surface. The intermediate species are adsorbed on A and B cation sites of each termination for both surfaces, then the free energy change of each intermediate reaction step is calculated. The detailed reaction mechanisms used here can be found in the supporting information section S2. Since the electronic structure of O$_2$ (triplet ground state) is not well described by DFT, an experimental value of 4.92 eV is used in step D, which is the free energy change of the total reaction 2H$_2$O $\longrightarrow$ O$_2$ + 2H$_2$. We calculated the ZPE for each intermediate on the (001) surface with TaON termination. The values found in this work are only slightly different from the ones calculated for different materials \cite{Valdes2008}, which means that the ZPE does not vary significantly between surfaces and even materials. Consequently, the same ZPE shown in supporting information Tab. S2 are used for all terminations of both materials in this work. The differences ($\Delta$ZPE - T$\Delta$S) for each reaction step of all mechanisms are shown in the supporting information Tab. S3.

\subsubsection{Reactions at the B site}

\begin{figure} 
\centering
 \includegraphics[width=\columnwidth]{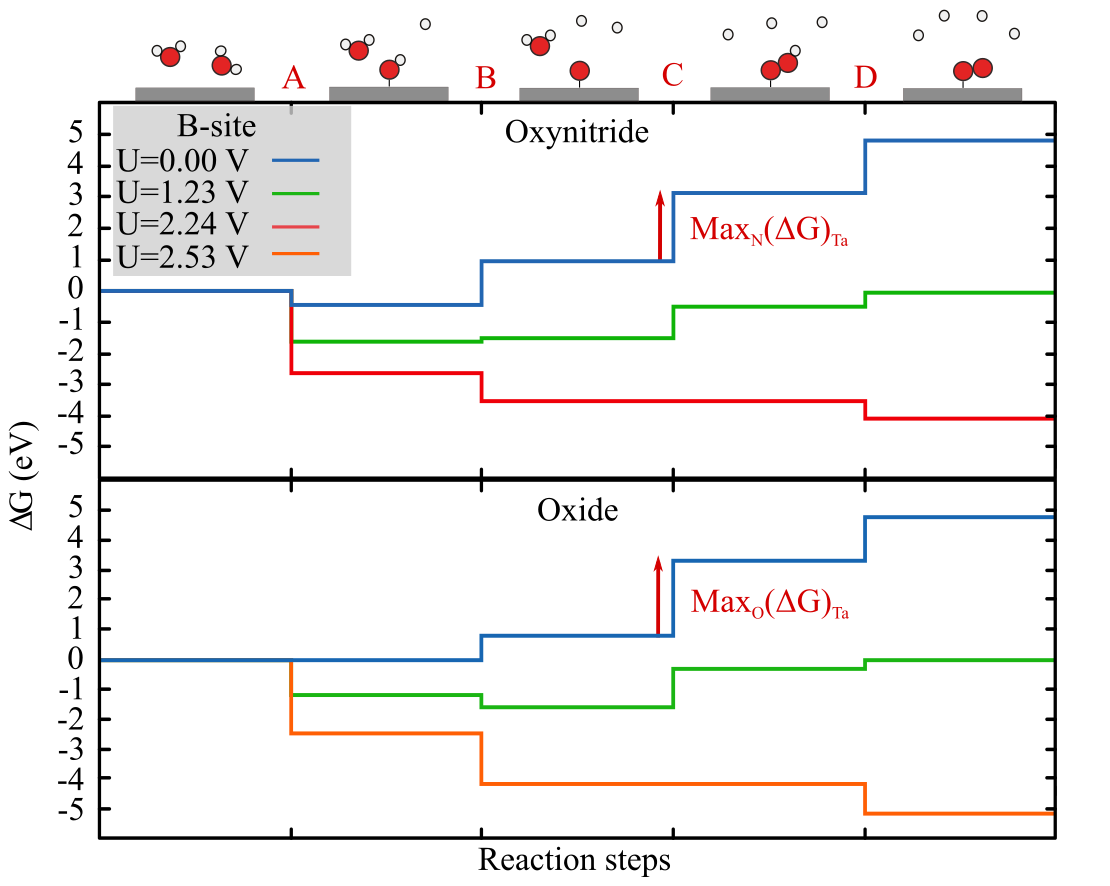}
\caption{Free energy profiles of the OER reaction steps at the B (Ta) site of the NaTaO$_3$ (113) and SrTaO$_2$N (001) surfaces at pH=0.}
\label{fig:FEDB}
\end{figure}

Fig. 5 shows the free energy diagram of the OER on the oxide and TaON-terminated oxynitride surface. The height of the steps corresponds to the free energy difference of the different reaction steps at various potentials U$_b$. At U$_b$ = 0 V all the steps are uphill except for the first step (* $\rightarrow$ *OH) on the oxide surface, which is already downhill in free energy. Step C (*O $\rightarrow$ *OOH), where the hydroperoxyl group is formed, has the highest free energy change for both surfaces (Max$\Delta$G) and thus represents the limiting step. At U$_b$ = 1.23 V, which is the standard equilibrium potential, some steps start to become downhill - A and B for the oxide and only A in the case of oxynitride - while the remaining steps are still uphill for both surfaces. As mentioned previously, step C has the highest free energy change for both surfaces at U$_b$ = 0 V and pH = 0. Since the height of this step is smaller on the oxynitride than the oxide surface (Fig. \ref{fig:FEDB}), the overpotential needed for water oxidation on the oxynitride (1.02 V) is smaller than for the oxide (1.30 V). Consequently, the Ta site on the oxynitride is predicted to be more active for the OER than on the oxide, which correlates with the general descriptor, where $\Delta$G$_{O}$-$\Delta$G$_{OH}$ is larger for the oxynitride than for the oxide.

\subsubsection{Reactions at the A site}

\begin{figure} 
\centering
 \includegraphics[width=\columnwidth]{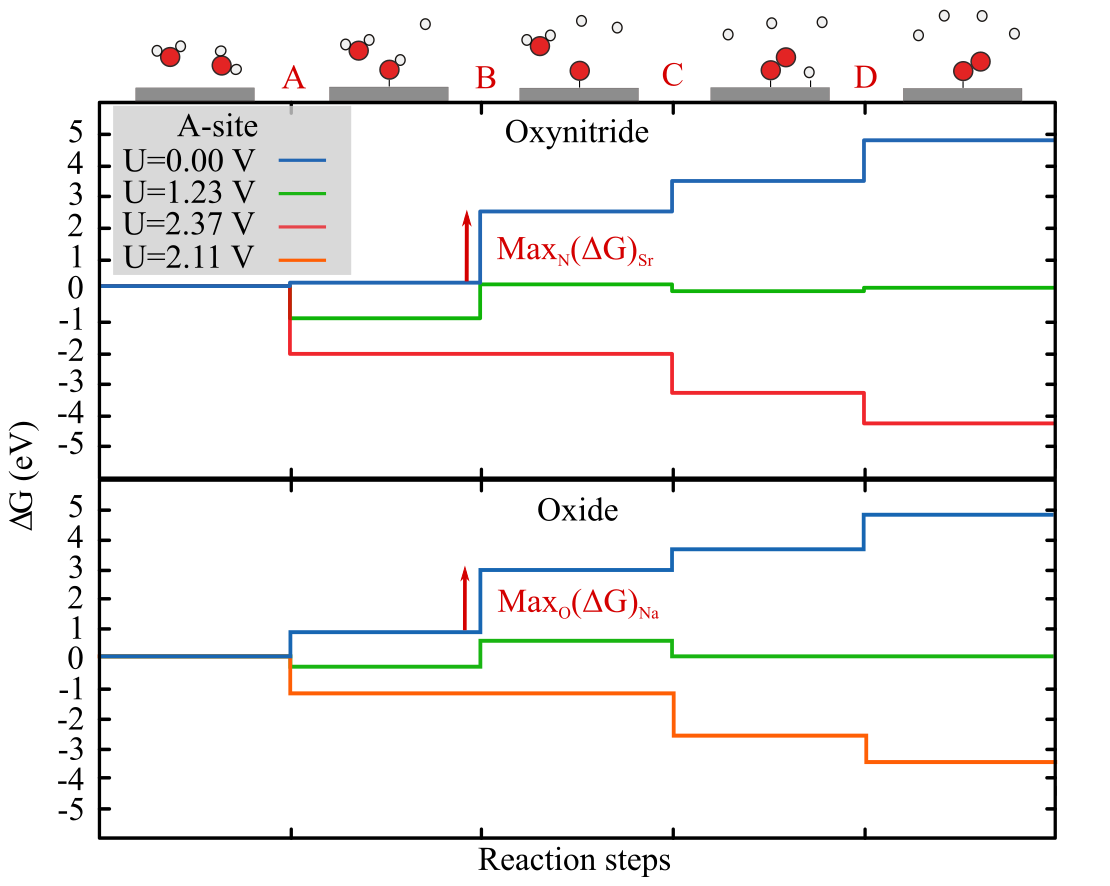}
\caption{Free energy profiles of the OER reaction steps at the A (Na, Sr) site of the NaTaO$_3$ (113) and SrTaO$_2$N (001) surfaces at pH=0.}
\label{fig:FEDA}
\end{figure}

The free energy diagram in Fig. \ref{fig:FEDA} reveals that at U = 0 V all the reaction steps on the A (Na, Sr) site are uphill for both materials, step A (*OH formation) on the oxynitride surface being almost flat. It is interesting here to mention that the hydroperoxyl group *OOH is not stable, the proton being transferred to a neighbouring lattice oxygen. We observed this to occur spontaneously on the oxide surface while on the oxynitride surface both the molecular and dissociated *OOH configurations have the same energy and proton transfer is not spontaneous. Following the mechanism suggested by Halck \etal \cite{Halck2014}, we compute the cycle by removing the proton bound to lattice oxygen. Contrary to what we found for the Ta site, step B is rate limiting on both the oxide and the oxynitride and its energy difference is larger on the oxynitride than on the oxide. An overpotential of $\eta$ = 1.14 and 0.89 V must be applied to make all the reactions downhill in free energy for the oxynitride and the oxide respectively. Therefore, the Na site on the oxide (113) surface, is more active than the Ta site, while on the oxynitride (001) surface the Ta site is the most active.

\subsection{OER mechanisms under operating conditions}

In this section, Pourbaix diagrams are computed for each surface termination of the oxide and the oxynitride to predict the most relevant surface coverage under (photo)-electrocatalytic operating conditions. In this goal, we calculate the free energy of each surface covered with $\nicefrac{1}{4}$ ML to 1 ML of the most stable intermediates *O and *OH, which form from adsorbed water after release of one proton and one electron or two protons and two electrons respectively. In the case of the oxynitride (001) surface there are either four Sr or four Ta adsorption sites on each termination, while on the oxide (113) surface eight Na and four 4 Ta sites are exposed at the same time. On the latter surface we therefore initially explored covering the two kinds of sites separately before considering mixed configurations involving both adsorption sites. 
 
\begin{figure}
 \includegraphics[width=\columnwidth]{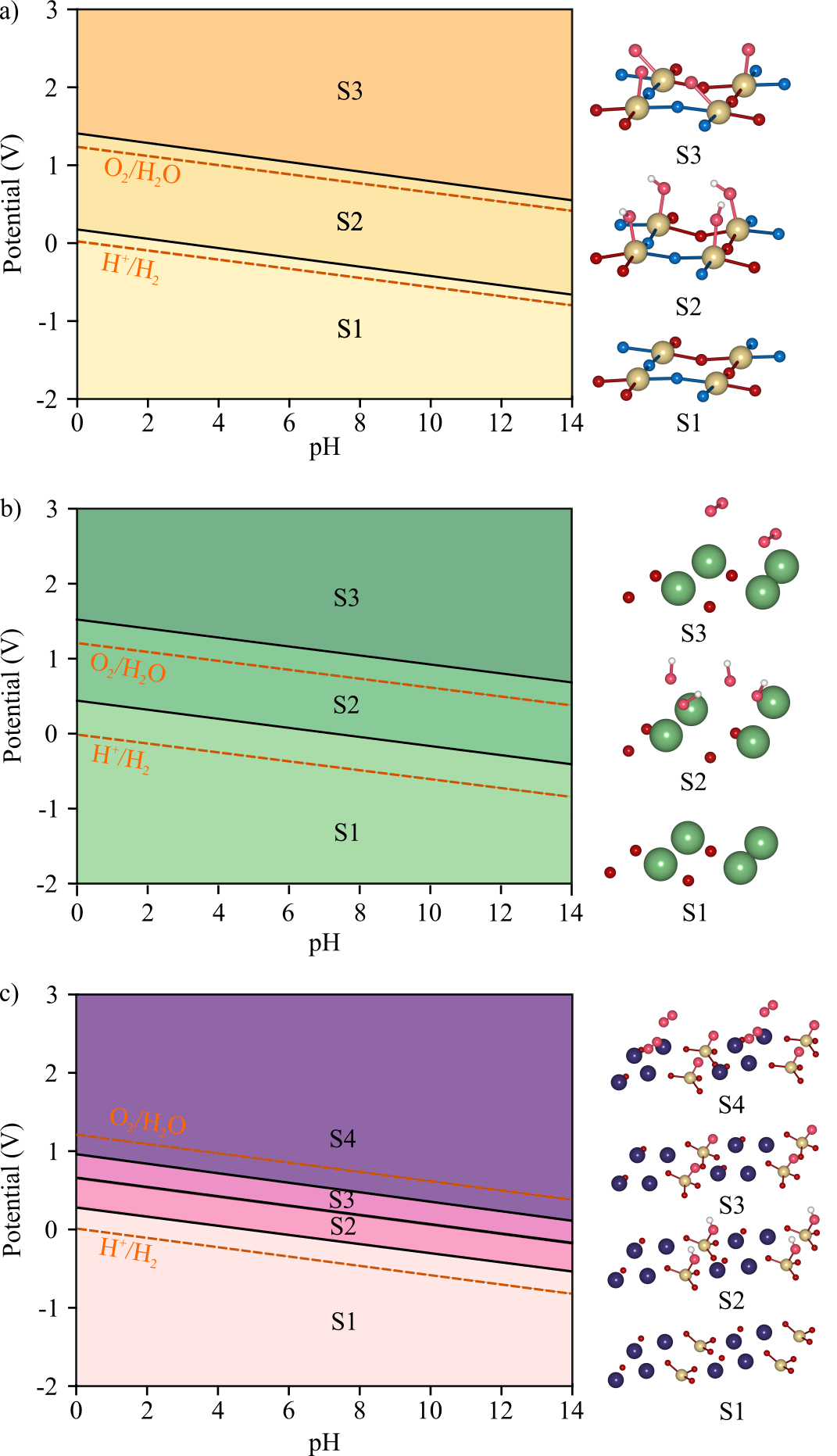}
\caption{Computed Pourbaix diagrams of a) the TaON-terminated SrTaO$_2$N (001), b) the SrO-terminated SrTaO$_2$N (001) and c) the NaTaO$_3$ (113) surface, pink and white atoms are the adsorbed oxygen and hydrogen atoms respectively.}
\label{fig:Pourbaix}
\end{figure}

As shown in Fig. \ref{fig:Pourbaix} the bare surfaces of both the oxide and the oxynitride are only stable at very low potential. As the potential increases, phase transitions are observed, in which the surfaces become covered with *OH adsorbates. For the oxynitride TaON termination at pH 0, this transition happens at the lowest potential ($\sim$0.17 V, followed by the oxide at $\sim$0.25 V and the oxynitride SrO termination at $\sim$0.44 V. The oxynitride surfaces stay in this hydroxylated state for potentials up to $\sim$1.52 V and $\sim$1.37 V for the SrO and TaON terminated surfaces respectively, whereas on the oxide surface *O adsorbates at the Ta site already start to appear at lower potentials of $\sim$0.66 V.

We investigated the surface adsorbate structures at potentials higher than 1.23 V and found the oxygen adsorbates on the TaON terminated oxynitride surface to occupy the bridge site above Ta\textendash N bonds (see sketch S3 in Fig. \ref{fig:Pourbaix} a). Despite this slight structural alteration, water oxidation proceed by the conventional mechanism starting with the *OOH formation step. We find however that the *OOH decays into a *O and a *OH, both adsorbed on Ta, the reaction proceeding by mechanism 2 as described in the supporting information section S2.b. We find that an overpotential of 0.88 V (supporting information Tab. S4) is needed to drive this reaction under electrochemical conditions, which is lower than the one calculated above (1.01 V) by considering the bare surface. 

As shown in Fig. \ref{fig:Pourbaix} b and c we find that *O adsorbed on Sr and Na sites spontaneously combine to form O$_2$, which then desorbs. No such *O coupling is observed at the Ta site for both the oxide and the oxynitride, due to the much higher adsorption energy of oxygen on Ta compared to Sr and Na. Additionally, on the oxide surface, the distance between Ta sites is 5.47 $\angstrom$ which hinders O$_2$ formation by this mechanism. In light of these observations O$_2$ could form by direct coupling of neighbouring adsorbed oxygen atoms on the SrO termination of the oxynitride or on the oxide surface (mechanism 3 in the supporting information section S2.b). For the SrO termination of the oxynitride we predict an overpotential of 2.17 V for mechanism 3 (supporting information Tab. S5), which is much higher than for the conventional mechanism 1 on a surface from which all *O adsorbates have desorbed by coupling. We hence predict that even under operating conditions the OER will proceed by mechanism 1 on this surface and have an overpotential of 1.14 V. We want to note however that the SrO termination is known to dissolve in aqueous solution \cite{Ohzuku2016}, which could render reactions on the SrO termination less relevant for photocatalysis.

For the oxide surface, the situation is more complex as after O$_2$ desorption by coupling of *O at Na sites, the Ta sites are still covered by *O. There could hence be three possible competing mechanisms. In the first mechanism the OER proceeds by the conventional mechanism on the Ta site, starting from step C (*OOH formation). This mechanism has an overpotential of 0.93 V (see supporting information Tab. S6), which is smaller than what we reported above for the uncovered surface (1.30 V), in agreement with the usual effect of adsorbate coverage \cite{Montoya2015a}. For the remaining mechanisms, we consider, similar to the oxynitride SrO termination, the conventional mechanism at a Na site, however in presence of *O adsorbates at the Ta sites, as well as the coupling of *O from neighbouring Na sites. The first of these mechanisms is predicted to have an overpotential of 0.99 V (supporting information Tab. S6), which is higher than the overpotential of 0.88 V reported above for the surface without *O at Ta sites. The coupling mechanism \cite{Lodi1978} on the other hand has the lowest overpotential of 0.79 V (supporting information Tab. S6) and is likely to be the active OER mechanism on the oxide surface under typical operating conditions.

\section{Conclusion}

We have investigated the oxygen evolution reaction on the NaTaO$_3$ (113) surface, which we find to be non-polar, and the SrTaO$_2$N (001) surface by using DFT calculations. On the clean oxynitride surface we find the Ta site to be most active for the OER with an overpotential of 1.02 V. On the clean oxide surface on the other hand, we find the lowest OER overpotential of 0.89 V for the Na site. Our calculations do unfortunately not allow us to ascertain if this difference stems from the different chemical composition of the surface or from the different surface orientations.

Under (photo)electrochemical conditions we have shown through computed surface Pourbaix diagrams that water oxidation takes place on oxygen covered surfaces. On the TaON terminated oxynitride surface this coverage effect lowers the overpotential to 0.88 V. We however find that neighbouring *O adsorbed to either Na or Sr couple and desorb as O$_2$, leaving these sites empty under operating conditions. Consequently we consider alternative OER mechanisms under these (photo)electrochemical conditions and show that on the SrO terminated oxynitride surface the reaction proceeds by the conventional *OOH mechanism with an overpotential of 1.14 V. For the oxide surface with *O-covered Ta on the other hand, a direct coupling of *O at neighbouring Na sites has the lowest overpotential of 0.79 V.

These calculations imply that the oxide (lowest overpotential 0.79 V) is a more active OER catalyst than the oxynitride (lowest overpotential of 0.88 V), however due to the influence of the A site rather than the B site as one would expect. Given the improved light absorption of the oxynitride, this will lead to a tradeoff between catalytic activity and light absorption for these two classes of materials.

\section{Acknowledgements}

This work is achieved thanks the SNF Professorship Grant PP00P2-157615. The calculations were performed on UBELIX (http://www.id.unibe.ch/hpc), the HPC cluster at the University of Bern.

\section{References}
\bibliographystyle{elsarticle-num.bst}
\bibliography{library}

\begin{thebibliography}{41}%
\makeatletter
\providecommand \@ifxundefined [1]{%
 \@ifx{#1\undefined}
}%
\providecommand \@ifnum [1]{%
 \ifnum #1\expandafter \@firstoftwo
 \else \expandafter \@secondoftwo
 \fi
}%
\providecommand \@ifx [1]{%
 \ifx #1\expandafter \@firstoftwo
 \else \expandafter \@secondoftwo
 \fi
}%
\providecommand \natexlab [1]{#1}%
\providecommand \enquote  [1]{``#1''}%
\providecommand \bibnamefont  [1]{#1}%
\providecommand \bibfnamefont [1]{#1}%
\providecommand \citenamefont [1]{#1}%
\providecommand \href@noop [0]{\@secondoftwo}%
\providecommand \href [0]{\begingroup \@sanitize@url \@href}%
\providecommand \@href[1]{\@@startlink{#1}\@@href}%
\providecommand \@@href[1]{\endgroup#1\@@endlink}%
\providecommand \@sanitize@url [0]{\catcode `\\12\catcode `\$12\catcode
  `\&12\catcode `\#12\catcode `\^12\catcode `\_12\catcode `\%12\relax}%
\providecommand \@@startlink[1]{}%
\providecommand \@@endlink[0]{}%
\providecommand \url  [0]{\begingroup\@sanitize@url \@url }%
\providecommand \@url [1]{\endgroup\@href {#1}{\urlprefix }}%
\providecommand \urlprefix  [0]{URL }%
\providecommand \Eprint [0]{\href }%
\providecommand \doibase [0]{http://dx.doi.org/}%
\providecommand \selectlanguage [0]{\@gobble}%
\providecommand \bibinfo  [0]{\@secondoftwo}%
\providecommand \bibfield  [0]{\@secondoftwo}%
\providecommand \translation [1]{[#1]}%
\providecommand \BibitemOpen [0]{}%
\providecommand \bibitemStop [0]{}%
\providecommand \bibitemNoStop [0]{.\EOS\space}%
\providecommand \EOS [0]{\spacefactor3000\relax}%
\providecommand \BibitemShut  [1]{\csname bibitem#1\endcsname}%
\let\auto@bib@innerbib\@empty
\bibitem [{\citenamefont {Fujishima}\ and\ \citenamefont
  {Honda}(1972)}]{Fujishima}%
  \BibitemOpen
  \bibfield  {author} {\bibinfo {author} {\bibfnamefont {A.}~\bibnamefont
  {Fujishima}}\ and\ \bibinfo {author} {\bibfnamefont {K.}~\bibnamefont
  {Honda}},\ }\href {\doibase 10.1038/238037a0} {\bibfield  {journal} {\bibinfo
   {journal} {Nature}\ }\textbf {\bibinfo {volume} {238}},\ \bibinfo {pages}
  {37} (\bibinfo {year} {1972})}\BibitemShut {NoStop}%
\bibitem [{\citenamefont {Osterloh}(2008)}]{Osterloh2008}%
  \BibitemOpen
  \bibfield  {author} {\bibinfo {author} {\bibfnamefont {F.~E.}\ \bibnamefont
  {Osterloh}},\ }\href {\doibase 10.1021/cm7024203} {\bibfield  {journal}
  {\bibinfo  {journal} {Chem. Mater.}\ }\textbf {\bibinfo {volume} {20}},\
  \bibinfo {pages} {35} (\bibinfo {year} {2008})}\BibitemShut {NoStop}%
\bibitem [{\citenamefont {Grabowska}(2016)}]{Grabowska2016}%
  \BibitemOpen
  \bibfield  {author} {\bibinfo {author} {\bibfnamefont {E.}~\bibnamefont
  {Grabowska}},\ }\href {\doibase 10.1016/j.apcatb.2015.12.035} {\bibfield
  {journal} {\bibinfo  {journal} {Applied Catalysis B: Environmental}\ }\textbf
  {\bibinfo {volume} {186}},\ \bibinfo {pages} {97} (\bibinfo {year}
  {2016})}\BibitemShut {NoStop}%
\bibitem [{\citenamefont {Modak}\ and\ \citenamefont
  {Ghosh}(2016)}]{Modak2016}%
  \BibitemOpen
  \bibfield  {author} {\bibinfo {author} {\bibfnamefont {B.}~\bibnamefont
  {Modak}}\ and\ \bibinfo {author} {\bibfnamefont {S.~K.}\ \bibnamefont
  {Ghosh}},\ }\href {\doibase 10.1021/acs.jpcc.5b11777} {\bibfield  {journal}
  {\bibinfo  {journal} {The Journal of Physical Chemistry C}\ }\textbf
  {\bibinfo {volume} {120}},\ \bibinfo {pages} {6920} (\bibinfo {year}
  {2016})}\BibitemShut {NoStop}%
\bibitem [{\citenamefont {Takata}\ \emph {et~al.}(2016)\citenamefont {Takata},
  \citenamefont {Pan},\ and\ \citenamefont {Domen}}]{Takata2016}%
  \BibitemOpen
  \bibfield  {author} {\bibinfo {author} {\bibfnamefont {T.}~\bibnamefont
  {Takata}}, \bibinfo {author} {\bibfnamefont {C.}~\bibnamefont {Pan}}, \ and\
  \bibinfo {author} {\bibfnamefont {K.}~\bibnamefont {Domen}},\ }\href
  {\doibase 10.1002/celc.201500324} {\bibfield  {journal} {\bibinfo  {journal}
  {ChemElectroChem}\ }\textbf {\bibinfo {volume} {3}},\ \bibinfo {pages} {31}
  (\bibinfo {year} {2016})}\BibitemShut {NoStop}%
\bibitem [{\citenamefont {Castelli}\ \emph {et~al.}(2012)\citenamefont
  {Castelli}, \citenamefont {Olsen}, \citenamefont {Datta}, \citenamefont
  {Landis}, \citenamefont {Dahl}, \citenamefont {Thygesen},\ and\ \citenamefont
  {Jacobsen}}]{Castelli2012}%
  \BibitemOpen
  \bibfield  {author} {\bibinfo {author} {\bibfnamefont {I.~E.}\ \bibnamefont
  {Castelli}}, \bibinfo {author} {\bibfnamefont {T.}~\bibnamefont {Olsen}},
  \bibinfo {author} {\bibfnamefont {S.}~\bibnamefont {Datta}}, \bibinfo
  {author} {\bibfnamefont {D.~D.}\ \bibnamefont {Landis}}, \bibinfo {author}
  {\bibfnamefont {S.}~\bibnamefont {Dahl}}, \bibinfo {author} {\bibfnamefont
  {K.~S.}\ \bibnamefont {Thygesen}}, \ and\ \bibinfo {author} {\bibfnamefont
  {K.~W.}\ \bibnamefont {Jacobsen}},\ }\href {\doibase 10.1039/C1EE02717D}
  {\bibfield  {journal} {\bibinfo  {journal} {Energy Environ. Sci.}\ }\textbf
  {\bibinfo {volume} {5}},\ \bibinfo {pages} {5814} (\bibinfo {year}
  {2012})}\BibitemShut {NoStop}%
\bibitem [{\citenamefont {Wu}\ \emph {et~al.}(2013)\citenamefont {Wu},
  \citenamefont {Lazic}, \citenamefont {Hautier}, \citenamefont {Persson},\
  and\ \citenamefont {Ceder}}]{Wu2013}%
  \BibitemOpen
  \bibfield  {author} {\bibinfo {author} {\bibfnamefont {Y.}~\bibnamefont
  {Wu}}, \bibinfo {author} {\bibfnamefont {P.}~\bibnamefont {Lazic}}, \bibinfo
  {author} {\bibfnamefont {G.}~\bibnamefont {Hautier}}, \bibinfo {author}
  {\bibfnamefont {K.}~\bibnamefont {Persson}}, \ and\ \bibinfo {author}
  {\bibfnamefont {G.}~\bibnamefont {Ceder}},\ }\href {\doibase
  10.1039/C2EE23482C} {\bibfield  {journal} {\bibinfo  {journal} {Energy
  Environ. Sci.}\ }\textbf {\bibinfo {volume} {6}},\ \bibinfo {pages} {157}
  (\bibinfo {year} {2013})}\BibitemShut {NoStop}%
\bibitem [{\citenamefont {Pe{\~{n}}a}\ and\ \citenamefont
  {Fierro}(2001)}]{Pena2001}%
  \BibitemOpen
  \bibfield  {author} {\bibinfo {author} {\bibfnamefont {M.~A.}\ \bibnamefont
  {Pe{\~{n}}a}}\ and\ \bibinfo {author} {\bibfnamefont {J.~L.~G.}\ \bibnamefont
  {Fierro}},\ }\href {\doibase 10.1021/cr980129f} {\bibfield  {journal}
  {\bibinfo  {journal} {Chemical Reviews}\ }\textbf {\bibinfo {volume} {101}},\
  \bibinfo {pages} {1981} (\bibinfo {year} {2001})}\BibitemShut {NoStop}%
\bibitem [{\citenamefont {Man}\ \emph {et~al.}(2011)\citenamefont {Man},
  \citenamefont {Su}, \citenamefont {Calle-Vallejo}, \citenamefont {Hansen},
  \citenamefont {Mart{\'{i}}nez}, \citenamefont {Inoglu}, \citenamefont
  {Kitchin}, \citenamefont {Jaramillo}, \citenamefont {N{\o}rskov},\ and\
  \citenamefont {Rossmeisl}}]{Man2011}%
  \BibitemOpen
  \bibfield  {author} {\bibinfo {author} {\bibfnamefont {I.~C.}\ \bibnamefont
  {Man}}, \bibinfo {author} {\bibfnamefont {H.~Y.}\ \bibnamefont {Su}},
  \bibinfo {author} {\bibfnamefont {F.}~\bibnamefont {Calle-Vallejo}}, \bibinfo
  {author} {\bibfnamefont {H.~A.}\ \bibnamefont {Hansen}}, \bibinfo {author}
  {\bibfnamefont {J.~I.}\ \bibnamefont {Mart{\'{i}}nez}}, \bibinfo {author}
  {\bibfnamefont {N.~G.}\ \bibnamefont {Inoglu}}, \bibinfo {author}
  {\bibfnamefont {J.}~\bibnamefont {Kitchin}}, \bibinfo {author} {\bibfnamefont
  {T.~F.}\ \bibnamefont {Jaramillo}}, \bibinfo {author} {\bibfnamefont {J.~K.}\
  \bibnamefont {N{\o}rskov}}, \ and\ \bibinfo {author} {\bibfnamefont
  {J.}~\bibnamefont {Rossmeisl}},\ }\href {\doibase 10.1002/cctc.201000397}
  {\bibfield  {journal} {\bibinfo  {journal} {ChemCatChem}\ }\textbf {\bibinfo
  {volume} {3}},\ \bibinfo {pages} {1159} (\bibinfo {year} {2011})}\BibitemShut
  {NoStop}%
\bibitem [{\citenamefont {Montoya}\ \emph {et~al.}(2015)\citenamefont
  {Montoya}, \citenamefont {Garcia-Mota}, \citenamefont {N{\o}rskov},\ and\
  \citenamefont {Vojvodic}}]{Montoya2015a}%
  \BibitemOpen
  \bibfield  {author} {\bibinfo {author} {\bibfnamefont {J.~H.}\ \bibnamefont
  {Montoya}}, \bibinfo {author} {\bibfnamefont {M.}~\bibnamefont
  {Garcia-Mota}}, \bibinfo {author} {\bibfnamefont {J.~K.}\ \bibnamefont
  {N{\o}rskov}}, \ and\ \bibinfo {author} {\bibfnamefont {A.}~\bibnamefont
  {Vojvodic}},\ }\href {\doibase 10.1039/c4cp05259e} {\bibfield  {journal}
  {\bibinfo  {journal} {Physical chemistry chemical physics : PCCP}\ }\textbf
  {\bibinfo {volume} {17}},\ \bibinfo {pages} {2634} (\bibinfo {year}
  {2015})}\BibitemShut {NoStop}%
\bibitem [{\citenamefont {Montoya}\ \emph {et~al.}(2018)\citenamefont
  {Montoya}, \citenamefont {Doyle}, \citenamefont {N{\o}rskov},\ and\
  \citenamefont {Vojvodic}}]{Montoya2018}%
  \BibitemOpen
  \bibfield  {author} {\bibinfo {author} {\bibfnamefont {J.~H.}\ \bibnamefont
  {Montoya}}, \bibinfo {author} {\bibfnamefont {A.~D.}\ \bibnamefont {Doyle}},
  \bibinfo {author} {\bibfnamefont {J.~K.}\ \bibnamefont {N{\o}rskov}}, \ and\
  \bibinfo {author} {\bibfnamefont {A.}~\bibnamefont {Vojvodic}},\ }\href
  {\doibase 10.1039/C7CP06539F} {\bibfield  {journal} {\bibinfo  {journal}
  {Physical Chemistry Chemical Physics}\ }\textbf {\bibinfo {volume} {20}},\
  \bibinfo {pages} {3813} (\bibinfo {year} {2018})}\BibitemShut {NoStop}%
\bibitem [{\citenamefont {Whitesides}\ and\ \citenamefont
  {Crabtree}(2007)}]{Whitesides2007}%
  \BibitemOpen
  \bibfield  {author} {\bibinfo {author} {\bibfnamefont {G.~M.}\ \bibnamefont
  {Whitesides}}\ and\ \bibinfo {author} {\bibfnamefont {G.~W.}\ \bibnamefont
  {Crabtree}},\ }\href {\doibase 10.1126/science.1140362} {\bibfield  {journal}
  {\bibinfo  {journal} {Science}\ }\textbf {\bibinfo {volume} {315}},\ \bibinfo
  {pages} {796} (\bibinfo {year} {2007})}\BibitemShut {NoStop}%
\bibitem [{\citenamefont {Hong}\ \emph {et~al.}(2015)\citenamefont {Hong},
  \citenamefont {Risch}, \citenamefont {Stoerzinger}, \citenamefont {Grimaud},
  \citenamefont {Suntivich},\ and\ \citenamefont {Shao-Horn}}]{Hong2015}%
  \BibitemOpen
  \bibfield  {author} {\bibinfo {author} {\bibfnamefont {W.~T.}\ \bibnamefont
  {Hong}}, \bibinfo {author} {\bibfnamefont {M.}~\bibnamefont {Risch}},
  \bibinfo {author} {\bibfnamefont {K.~A.}\ \bibnamefont {Stoerzinger}},
  \bibinfo {author} {\bibfnamefont {A.}~\bibnamefont {Grimaud}}, \bibinfo
  {author} {\bibfnamefont {J.}~\bibnamefont {Suntivich}}, \ and\ \bibinfo
  {author} {\bibfnamefont {Y.}~\bibnamefont {Shao-Horn}},\ }\href {\doibase
  10.1039/C4EE03869J} {\bibfield  {journal} {\bibinfo  {journal} {Energy
  Environ. Sci.}\ }\textbf {\bibinfo {volume} {8}},\ \bibinfo {pages} {1404}
  (\bibinfo {year} {2015})}\BibitemShut {NoStop}%
\bibitem [{\citenamefont {Zhang}\ and\ \citenamefont
  {Bieberle-H{\"{u}}tter}(2016)}]{Zhang2016a}%
  \BibitemOpen
  \bibfield  {author} {\bibinfo {author} {\bibfnamefont {X.}~\bibnamefont
  {Zhang}}\ and\ \bibinfo {author} {\bibfnamefont {A.}~\bibnamefont
  {Bieberle-H{\"{u}}tter}},\ }\href {\doibase 10.1002/cssc.201600214}
  {\bibfield  {journal} {\bibinfo  {journal} {ChemSusChem}\ }\textbf {\bibinfo
  {volume} {9}},\ \bibinfo {pages} {1223} (\bibinfo {year} {2016})}\BibitemShut
  {NoStop}%
\bibitem [{\citenamefont {N{\o}rskov}\ \emph {et~al.}(2004)\citenamefont
  {N{\o}rskov}, \citenamefont {Ressmeisl}, \citenamefont {Logadottir},
  \citenamefont {Lindqvist}, \citenamefont {Kitchin}, \citenamefont
  {Bligaard},\ and\ \citenamefont {Jonsson}}]{Norskov2004}%
  \BibitemOpen
  \bibfield  {author} {\bibinfo {author} {\bibfnamefont {J.~K.}\ \bibnamefont
  {N{\o}rskov}}, \bibinfo {author} {\bibfnamefont {J.}~\bibnamefont
  {Ressmeisl}}, \bibinfo {author} {\bibfnamefont {A.}~\bibnamefont
  {Logadottir}}, \bibinfo {author} {\bibfnamefont {L.}~\bibnamefont
  {Lindqvist}}, \bibinfo {author} {\bibfnamefont {J.~R.}\ \bibnamefont
  {Kitchin}}, \bibinfo {author} {\bibfnamefont {T.}~\bibnamefont {Bligaard}}, \
  and\ \bibinfo {author} {\bibfnamefont {H.}~\bibnamefont {Jonsson}},\ }\href
  {\doibase 10.1021/jp047349j} {\bibfield  {journal} {\bibinfo  {journal} {J.
  Phys. Chem. B}\ }\textbf {\bibinfo {volume} {108}},\ \bibinfo {pages} {17886}
  (\bibinfo {year} {2004})}\BibitemShut {NoStop}%
\bibitem [{\citenamefont {Rossmeisl}\ \emph {et~al.}(2007)\citenamefont
  {Rossmeisl}, \citenamefont {Qu}, \citenamefont {Zhu}, \citenamefont {Kroes},\
  and\ \citenamefont {N{\o}rskov}}]{Rossmeisl2007}%
  \BibitemOpen
  \bibfield  {author} {\bibinfo {author} {\bibfnamefont {J.}~\bibnamefont
  {Rossmeisl}}, \bibinfo {author} {\bibfnamefont {Z.~W.}\ \bibnamefont {Qu}},
  \bibinfo {author} {\bibfnamefont {H.}~\bibnamefont {Zhu}}, \bibinfo {author}
  {\bibfnamefont {G.~J.}\ \bibnamefont {Kroes}}, \ and\ \bibinfo {author}
  {\bibfnamefont {J.~K.}\ \bibnamefont {N{\o}rskov}},\ }\href {\doibase
  10.1016/j.jelechem.2006.11.008} {\bibfield  {journal} {\bibinfo  {journal}
  {Journal of Electroanalytical Chemistry}\ }\textbf {\bibinfo {volume}
  {607}},\ \bibinfo {pages} {83} (\bibinfo {year} {2007})}\BibitemShut
  {NoStop}%
\bibitem [{\citenamefont {Nguyen}\ \emph {et~al.}(2015)\citenamefont {Nguyen},
  \citenamefont {Piccinin}, \citenamefont {Seriani},\ and\ \citenamefont
  {Gebauer}}]{Nguyen2015}%
  \BibitemOpen
  \bibfield  {author} {\bibinfo {author} {\bibfnamefont {M.~T.}\ \bibnamefont
  {Nguyen}}, \bibinfo {author} {\bibfnamefont {S.}~\bibnamefont {Piccinin}},
  \bibinfo {author} {\bibfnamefont {N.}~\bibnamefont {Seriani}}, \ and\
  \bibinfo {author} {\bibfnamefont {R.}~\bibnamefont {Gebauer}},\ }\href
  {\doibase 10.1021/cs5017326} {\bibfield  {journal} {\bibinfo  {journal} {ACS
  Catalysis}\ }\textbf {\bibinfo {volume} {5}},\ \bibinfo {pages} {715}
  (\bibinfo {year} {2015})}\BibitemShut {NoStop}%
\bibitem [{\citenamefont {Trasatti}(1972)}]{Trasatti1972}%
  \BibitemOpen
  \bibfield  {author} {\bibinfo {author} {\bibfnamefont {S.}~\bibnamefont
  {Trasatti}},\ }\href {\doibase 10.1016/S0022-0728(72)80485-6} {\bibfield
  {journal} {\bibinfo  {journal} {Journal of Electroanalytical Chemistry}\
  }\textbf {\bibinfo {volume} {39}},\ \bibinfo {pages} {163} (\bibinfo {year}
  {1972})}\BibitemShut {NoStop}%
\bibitem [{\citenamefont {Russell}\ \emph {et~al.}(2008)\citenamefont
  {Russell}, \citenamefont {Shen}, \citenamefont {Tr{\"{a}}uble}, \citenamefont
  {Wittstock}, \citenamefont {Wasileski}, \citenamefont {Janik},\ and\
  \citenamefont {Chem}}]{Russell2008}%
  \BibitemOpen
  \bibfield  {author} {\bibinfo {author} {\bibfnamefont {A.}~\bibnamefont
  {Russell}}, \bibinfo {author} {\bibfnamefont {Y.}~\bibnamefont {Shen}},
  \bibinfo {author} {\bibfnamefont {M.}~\bibnamefont {Tr{\"{a}}uble}}, \bibinfo
  {author} {\bibfnamefont {G.}~\bibnamefont {Wittstock}}, \bibinfo {author}
  {\bibfnamefont {S.~A.}\ \bibnamefont {Wasileski}}, \bibinfo {author}
  {\bibfnamefont {M.~J.}\ \bibnamefont {Janik}}, \ and\ \bibinfo {author}
  {\bibfnamefont {P.}~\bibnamefont {Chem}},\ }\href {\doibase 10.1039/b808799g}
  {\bibfield  {journal} {\bibinfo  {journal} {Physical Chemistry Chemical
  Physics}\ }\textbf {\bibinfo {volume} {10}},\ \bibinfo {pages} {3607}
  (\bibinfo {year} {2008})}\BibitemShut {NoStop}%
\bibitem [{\citenamefont {Yamasita}\ \emph {et~al.}(2004)\citenamefont
  {Yamasita}, \citenamefont {Takata}, \citenamefont {Hara}, \citenamefont
  {Kondo},\ and\ \citenamefont {Domen}}]{Yamasita2004}%
  \BibitemOpen
  \bibfield  {author} {\bibinfo {author} {\bibfnamefont {D.}~\bibnamefont
  {Yamasita}}, \bibinfo {author} {\bibfnamefont {T.}~\bibnamefont {Takata}},
  \bibinfo {author} {\bibfnamefont {M.}~\bibnamefont {Hara}}, \bibinfo {author}
  {\bibfnamefont {J.~N.}\ \bibnamefont {Kondo}}, \ and\ \bibinfo {author}
  {\bibfnamefont {K.}~\bibnamefont {Domen}},\ }\href {\doibase
  10.1016/j.ssi.2004.04.033} {\bibfield  {journal} {\bibinfo  {journal} {Solid
  State Ionics}\ }\textbf {\bibinfo {volume} {172}},\ \bibinfo {pages} {591}
  (\bibinfo {year} {2004})}\BibitemShut {NoStop}%
\bibitem [{\citenamefont {Balaz}\ \emph {et~al.}(2013)\citenamefont {Balaz},
  \citenamefont {Porter}, \citenamefont {Woodward},\ and\ \citenamefont
  {Brillson}}]{Balaz2013}%
  \BibitemOpen
  \bibfield  {author} {\bibinfo {author} {\bibfnamefont {S.}~\bibnamefont
  {Balaz}}, \bibinfo {author} {\bibfnamefont {S.~H.}\ \bibnamefont {Porter}},
  \bibinfo {author} {\bibfnamefont {P.~M.}\ \bibnamefont {Woodward}}, \ and\
  \bibinfo {author} {\bibfnamefont {L.~J.}\ \bibnamefont {Brillson}},\ }\href
  {\doibase 10.1021/cm401815w} {\bibfield  {journal} {\bibinfo  {journal}
  {Chemistry of Materials}\ }\textbf {\bibinfo {volume} {25}},\ \bibinfo
  {pages} {3337} (\bibinfo {year} {2013})}\BibitemShut {NoStop}%
\bibitem [{\citenamefont {Giannozzi}\ \emph {et~al.}(2009)\citenamefont
  {Giannozzi}, \citenamefont {Baroni}, \citenamefont {Bonini}, \citenamefont
  {Calandra}, \citenamefont {Car}, \citenamefont {Cavazzoni}, \citenamefont
  {Ceresoli}, \citenamefont {Chiarotti}, \citenamefont {Cococcioni},
  \citenamefont {Dabo}, \citenamefont {Corso}, \citenamefont {Gironcoli},
  \citenamefont {Fabris}, \citenamefont {Fratesi}, \citenamefont {Gebauer},
  \citenamefont {Gerstmann}, \citenamefont {Gougoussis}, \citenamefont
  {Kokalj}, \citenamefont {Lazzeri}, \citenamefont {Martin-Samos},
  \citenamefont {Marzari}, \citenamefont {Mauri}, \citenamefont {Mazzarello},
  \citenamefont {Paolini}, \citenamefont {Pasquarello}, \citenamefont
  {Paulatto}, \citenamefont {Sbraccia}, \citenamefont {Scandolo}, \citenamefont
  {Sclauzero}, \citenamefont {Seitsonen}, \citenamefont {Smogunov},\ and\
  \citenamefont {Umari}}]{Giannozzi2009}%
  \BibitemOpen
  \bibfield  {author} {\bibinfo {author} {\bibfnamefont {P.}~\bibnamefont
  {Giannozzi}}, \bibinfo {author} {\bibfnamefont {S.}~\bibnamefont {Baroni}},
  \bibinfo {author} {\bibfnamefont {N.}~\bibnamefont {Bonini}}, \bibinfo
  {author} {\bibfnamefont {M.}~\bibnamefont {Calandra}}, \bibinfo {author}
  {\bibfnamefont {R.}~\bibnamefont {Car}}, \bibinfo {author} {\bibfnamefont
  {C.}~\bibnamefont {Cavazzoni}}, \bibinfo {author} {\bibfnamefont
  {D.}~\bibnamefont {Ceresoli}}, \bibinfo {author} {\bibfnamefont {G.~L.}\
  \bibnamefont {Chiarotti}}, \bibinfo {author} {\bibfnamefont {M.}~\bibnamefont
  {Cococcioni}}, \bibinfo {author} {\bibfnamefont {I.}~\bibnamefont {Dabo}},
  \bibinfo {author} {\bibfnamefont {A.~D.}\ \bibnamefont {Corso}}, \bibinfo
  {author} {\bibfnamefont {S.~D.}\ \bibnamefont {Gironcoli}}, \bibinfo {author}
  {\bibfnamefont {S.}~\bibnamefont {Fabris}}, \bibinfo {author} {\bibfnamefont
  {G.}~\bibnamefont {Fratesi}}, \bibinfo {author} {\bibfnamefont
  {R.}~\bibnamefont {Gebauer}}, \bibinfo {author} {\bibfnamefont
  {U.}~\bibnamefont {Gerstmann}}, \bibinfo {author} {\bibfnamefont
  {C.}~\bibnamefont {Gougoussis}}, \bibinfo {author} {\bibfnamefont
  {A.}~\bibnamefont {Kokalj}}, \bibinfo {author} {\bibfnamefont
  {M.}~\bibnamefont {Lazzeri}}, \bibinfo {author} {\bibfnamefont
  {L.}~\bibnamefont {Martin-Samos}}, \bibinfo {author} {\bibfnamefont
  {N.}~\bibnamefont {Marzari}}, \bibinfo {author} {\bibfnamefont
  {F.}~\bibnamefont {Mauri}}, \bibinfo {author} {\bibfnamefont
  {R.}~\bibnamefont {Mazzarello}}, \bibinfo {author} {\bibfnamefont
  {S.}~\bibnamefont {Paolini}}, \bibinfo {author} {\bibfnamefont
  {A.}~\bibnamefont {Pasquarello}}, \bibinfo {author} {\bibfnamefont
  {L.}~\bibnamefont {Paulatto}}, \bibinfo {author} {\bibfnamefont
  {C.}~\bibnamefont {Sbraccia}}, \bibinfo {author} {\bibfnamefont
  {S.}~\bibnamefont {Scandolo}}, \bibinfo {author} {\bibfnamefont
  {G.}~\bibnamefont {Sclauzero}}, \bibinfo {author} {\bibfnamefont {A.~P.}\
  \bibnamefont {Seitsonen}}, \bibinfo {author} {\bibfnamefont {A.}~\bibnamefont
  {Smogunov}}, \ and\ \bibinfo {author} {\bibfnamefont {P.}~\bibnamefont
  {Umari}},\ }\href@noop {} {\bibfield  {journal} {\bibinfo  {journal} {Journal
  of Physics: Condensed Matter}\ }\textbf {\bibinfo {volume} {21}},\ \bibinfo
  {pages} {395502} (\bibinfo {year} {2009})}\BibitemShut {NoStop}%
\bibitem [{\citenamefont {Perdew}\ \emph {et~al.}(1996)\citenamefont {Perdew},
  \citenamefont {Burke},\ and\ \citenamefont {Ernzerhof}}]{Perdew1996}%
  \BibitemOpen
  \bibfield  {author} {\bibinfo {author} {\bibfnamefont {J.~P.}\ \bibnamefont
  {Perdew}}, \bibinfo {author} {\bibfnamefont {K.}~\bibnamefont {Burke}}, \
  and\ \bibinfo {author} {\bibfnamefont {M.}~\bibnamefont {Ernzerhof}},\ }\href
  {\doibase 10.1103/PhysRevLett.77.3865} {\bibfield  {journal} {\bibinfo
  {journal} {Physical Review Letters}\ }\textbf {\bibinfo {volume} {77}},\
  \bibinfo {pages} {3865} (\bibinfo {year} {1996})}\BibitemShut {NoStop}%
\bibitem [{\citenamefont {Anisimov}\ \emph {et~al.}(1991)\citenamefont
  {Anisimov}, \citenamefont {Zaanen},\ and\ \citenamefont
  {Andersen}}]{Anisimov1991}%
  \BibitemOpen
  \bibfield  {author} {\bibinfo {author} {\bibfnamefont {V.~I.}\ \bibnamefont
  {Anisimov}}, \bibinfo {author} {\bibfnamefont {J.}~\bibnamefont {Zaanen}}, \
  and\ \bibinfo {author} {\bibfnamefont {O.~K.}\ \bibnamefont {Andersen}},\
  }\href {\doibase 10.1103/PhysRevB.44.943} {\bibfield  {journal} {\bibinfo
  {journal} {Physical Review B}\ }\textbf {\bibinfo {volume} {44}},\ \bibinfo
  {pages} {943} (\bibinfo {year} {1991})}\BibitemShut {NoStop}%
\bibitem [{\citenamefont {Vanderbilt}(1990)}]{Vanderbilt1990}%
  \BibitemOpen
  \bibfield  {author} {\bibinfo {author} {\bibfnamefont {D.}~\bibnamefont
  {Vanderbilt}},\ }\href {\doibase 10.1103/PhysRevB.41.7892} {\bibfield
  {journal} {\bibinfo  {journal} {Physical Review B}\ }\textbf {\bibinfo
  {volume} {41}},\ \bibinfo {pages} {7892} (\bibinfo {year}
  {1990})}\BibitemShut {NoStop}%
\bibitem [{\citenamefont {Ennedy}\ \emph {et~al.}(2006)\citenamefont {Ennedy},
  \citenamefont {Prodjosantoso},\ and\ \citenamefont {Howard}}]{Ennedy2006}%
  \BibitemOpen
  \bibfield  {author} {\bibinfo {author} {\bibfnamefont {B.~J.}\ \bibnamefont
  {Ennedy}}, \bibinfo {author} {\bibfnamefont {A.~K.}\ \bibnamefont
  {Prodjosantoso}}, \ and\ \bibinfo {author} {\bibfnamefont {C.~J.}\
  \bibnamefont {Howard}},\ }\href {\doibase 10.1088/0953-8984/11/33/302}
  {\bibfield  {journal} {\bibinfo  {journal} {Journal of Physics: Condensed
  Matter}\ }\textbf {\bibinfo {volume} {11}},\ \bibinfo {pages} {6319}
  (\bibinfo {year} {2006})}\BibitemShut {NoStop}%
\bibitem [{\citenamefont {Clarke}\ \emph {et~al.}(2002)\citenamefont {Clarke},
  \citenamefont {Hardstone}, \citenamefont {Michie},\ and\ \citenamefont
  {Rosseinsky}}]{Clarke2002}%
  \BibitemOpen
  \bibfield  {author} {\bibinfo {author} {\bibfnamefont {S.~J.}\ \bibnamefont
  {Clarke}}, \bibinfo {author} {\bibfnamefont {K.~A.}\ \bibnamefont
  {Hardstone}}, \bibinfo {author} {\bibfnamefont {C.~W.}\ \bibnamefont
  {Michie}}, \ and\ \bibinfo {author} {\bibfnamefont {M.~J.}\ \bibnamefont
  {Rosseinsky}},\ }\href {\doibase 10.1021/cm011738y} {\bibfield  {journal}
  {\bibinfo  {journal} {Chemistry of Materials}\ }\textbf {\bibinfo {volume}
  {14}},\ \bibinfo {pages} {2664} (\bibinfo {year} {2002})}\BibitemShut
  {NoStop}%
\bibitem [{\citenamefont {Pack}\ and\ \citenamefont
  {Monkhorst}(1977)}]{Pack1977}%
  \BibitemOpen
  \bibfield  {author} {\bibinfo {author} {\bibfnamefont {J.~D.}\ \bibnamefont
  {Pack}}\ and\ \bibinfo {author} {\bibfnamefont {H.~J.}\ \bibnamefont
  {Monkhorst}},\ }\href {\doibase 10.1103/PhysRevB.16.1748} {\bibfield
  {journal} {\bibinfo  {journal} {Physical Review B}\ }\textbf {\bibinfo
  {volume} {16}},\ \bibinfo {pages} {1748} (\bibinfo {year}
  {1977})}\BibitemShut {NoStop}%
\bibitem [{\citenamefont {Bengtsson}(1999)}]{Bengtsson1999}%
  \BibitemOpen
  \bibfield  {author} {\bibinfo {author} {\bibfnamefont {L.}~\bibnamefont
  {Bengtsson}},\ }\href {\doibase 10.1103/PhysRevB.59.12301} {\bibfield
  {journal} {\bibinfo  {journal} {Physical Review B}\ }\textbf {\bibinfo
  {volume} {59}},\ \bibinfo {pages} {12301} (\bibinfo {year}
  {1999})}\BibitemShut {NoStop}%
\bibitem [{\citenamefont {Togo}\ and\ \citenamefont
  {Tanaka}(2015)}]{Togo2015a}%
  \BibitemOpen
  \bibfield  {author} {\bibinfo {author} {\bibfnamefont {A.}~\bibnamefont
  {Togo}}\ and\ \bibinfo {author} {\bibfnamefont {I.}~\bibnamefont {Tanaka}},\
  }\href {\doibase 10.1016/j.scriptamat.2015.07.021} {\bibfield  {journal}
  {\bibinfo  {journal} {Scripta Materialia}\ }\textbf {\bibinfo {volume}
  {108}},\ \bibinfo {pages} {1} (\bibinfo {year} {2015})}\BibitemShut {NoStop}%
\bibitem [{\citenamefont {Chase}(1998)}]{Chase}%
  \BibitemOpen
  \bibfield  {author} {\bibinfo {author} {\bibfnamefont {M.~W.}\ \bibnamefont
  {Chase}},\ }\href@noop {} {\emph {\bibinfo {title} {{NIST-JANAF
  Thermochemical Tables}}}}\ (\bibinfo  {publisher} {American Chemical
  Society},\ \bibinfo {address} {New York},\ \bibinfo {year}
  {1998})\BibitemShut {NoStop}%
\bibitem [{\citenamefont {Clark}\ \emph {et~al.}(2013)\citenamefont {Clark},
  \citenamefont {Or{\'{o}}-Sol{\'{e}}}, \citenamefont {Knight}, \citenamefont
  {Fuertes},\ and\ \citenamefont {Attfield}}]{Clark2013}%
  \BibitemOpen
  \bibfield  {author} {\bibinfo {author} {\bibfnamefont {L.}~\bibnamefont
  {Clark}}, \bibinfo {author} {\bibfnamefont {J.}~\bibnamefont
  {Or{\'{o}}-Sol{\'{e}}}}, \bibinfo {author} {\bibfnamefont {K.~S.}\
  \bibnamefont {Knight}}, \bibinfo {author} {\bibfnamefont {A.}~\bibnamefont
  {Fuertes}}, \ and\ \bibinfo {author} {\bibfnamefont {J.~P.}\ \bibnamefont
  {Attfield}},\ }\href {\doibase 10.1021/cm4037132} {\bibfield  {journal}
  {\bibinfo  {journal} {Chemistry of Materials}\ }\textbf {\bibinfo {volume}
  {25}},\ \bibinfo {pages} {5004} (\bibinfo {year} {2013})}\BibitemShut
  {NoStop}%
\bibitem [{\citenamefont {Glazer}(1972)}]{Glazer:1972eb}%
  \BibitemOpen
  \bibfield  {author} {\bibinfo {author} {\bibfnamefont {A.~M.}\ \bibnamefont
  {Glazer}},\ }\href@noop {} {\bibfield  {journal} {\bibinfo  {journal} {Acta
  Cryst. B}\ }\textbf {\bibinfo {volume} {28}},\ \bibinfo {pages} {3384}
  (\bibinfo {year} {1972})}\BibitemShut {NoStop}%
\bibitem [{\citenamefont {Yang}\ \emph {et~al.}(2011)\citenamefont {Yang},
  \citenamefont {Or{\'{o}}-Sol{\'{e}}}, \citenamefont {Rodgers}, \citenamefont
  {Jorge}, \citenamefont {Fuertes},\ and\ \citenamefont
  {Attfield}}]{Yang2011b}%
  \BibitemOpen
  \bibfield  {author} {\bibinfo {author} {\bibfnamefont {M.}~\bibnamefont
  {Yang}}, \bibinfo {author} {\bibfnamefont {J.}~\bibnamefont
  {Or{\'{o}}-Sol{\'{e}}}}, \bibinfo {author} {\bibfnamefont {J.~a.}\
  \bibnamefont {Rodgers}}, \bibinfo {author} {\bibfnamefont {A.~B.}\
  \bibnamefont {Jorge}}, \bibinfo {author} {\bibfnamefont {A.}~\bibnamefont
  {Fuertes}}, \ and\ \bibinfo {author} {\bibfnamefont {J.~P.}\ \bibnamefont
  {Attfield}},\ }\href {\doibase 10.1038/nchem.908} {\bibfield  {journal}
  {\bibinfo  {journal} {Nature chemistry}\ }\textbf {\bibinfo {volume} {3}},\
  \bibinfo {pages} {47} (\bibinfo {year} {2011})}\BibitemShut {NoStop}%
\bibitem [{\citenamefont {Noguera}(2000)}]{Noguera2000}%
  \BibitemOpen
  \bibfield  {author} {\bibinfo {author} {\bibfnamefont {C.}~\bibnamefont
  {Noguera}},\ }\href {\doibase 10.1088/0953-8984/12/31/201} {\bibfield
  {journal} {\bibinfo  {journal} {Journal of Physics: Condensed Matter}\
  }\textbf {\bibinfo {volume} {12}},\ \bibinfo {pages} {R367} (\bibinfo {year}
  {2000})}\BibitemShut {NoStop}%
\bibitem [{\citenamefont {Noguera}\ and\ \citenamefont
  {Goniakowski}(2013)}]{Noguera2013}%
  \BibitemOpen
  \bibfield  {author} {\bibinfo {author} {\bibfnamefont {C.}~\bibnamefont
  {Noguera}}\ and\ \bibinfo {author} {\bibfnamefont {J.}~\bibnamefont
  {Goniakowski}},\ }\href {\doibase 10.1021/cr3003032} {\bibfield  {journal}
  {\bibinfo  {journal} {Chemical Reviews}\ }\textbf {\bibinfo {volume} {113}},\
  \bibinfo {pages} {4073} (\bibinfo {year} {2013})}\BibitemShut {NoStop}%
\bibitem [{\citenamefont {Deacon-Smith}\ \emph {et~al.}(2014)\citenamefont
  {Deacon-Smith}, \citenamefont {Scanlon}, \citenamefont {Catlow},
  \citenamefont {Sokol},\ and\ \citenamefont {Woodley}}]{Deacon-Smith2014a}%
  \BibitemOpen
  \bibfield  {author} {\bibinfo {author} {\bibfnamefont {D.~E.~E.}\
  \bibnamefont {Deacon-Smith}}, \bibinfo {author} {\bibfnamefont {D.~O.}\
  \bibnamefont {Scanlon}}, \bibinfo {author} {\bibfnamefont {C.~R.~A.}\
  \bibnamefont {Catlow}}, \bibinfo {author} {\bibfnamefont {A.~A.}\
  \bibnamefont {Sokol}}, \ and\ \bibinfo {author} {\bibfnamefont {S.~M.}\
  \bibnamefont {Woodley}},\ }\href {\doibase 10.1002/adma.201401858} {\bibfield
   {journal} {\bibinfo  {journal} {Advanced Materials}\ }\textbf {\bibinfo
  {volume} {26}},\ \bibinfo {pages} {7252} (\bibinfo {year}
  {2014})}\BibitemShut {NoStop}%
\bibitem [{\citenamefont {Vald{\'{e}}s}\ \emph {et~al.}(2008)\citenamefont
  {Vald{\'{e}}s}, \citenamefont {Qu}, \citenamefont {Kroes}, \citenamefont
  {Rossmeisl},\ and\ \citenamefont {N{\o}rskov}}]{Valdes2008}%
  \BibitemOpen
  \bibfield  {author} {\bibinfo {author} {\bibfnamefont {{\'{A}}.}~\bibnamefont
  {Vald{\'{e}}s}}, \bibinfo {author} {\bibfnamefont {Z.~W.}\ \bibnamefont
  {Qu}}, \bibinfo {author} {\bibfnamefont {G.~J.}\ \bibnamefont {Kroes}},
  \bibinfo {author} {\bibfnamefont {J.}~\bibnamefont {Rossmeisl}}, \ and\
  \bibinfo {author} {\bibfnamefont {J.~K.}\ \bibnamefont {N{\o}rskov}},\ }\href
  {\doibase 10.1021/jp711929d} {\bibfield  {journal} {\bibinfo  {journal}
  {Journal of Physical Chemistry C}\ }\textbf {\bibinfo {volume} {112}},\
  \bibinfo {pages} {9872} (\bibinfo {year} {2008})}\BibitemShut {NoStop}%
\bibitem [{\citenamefont {Halck}\ \emph {et~al.}(2014)\citenamefont {Halck},
  \citenamefont {Petrykin}, \citenamefont {Krtil},\ and\ \citenamefont
  {Rossmeisl}}]{Halck2014}%
  \BibitemOpen
  \bibfield  {author} {\bibinfo {author} {\bibfnamefont {N.~B.}\ \bibnamefont
  {Halck}}, \bibinfo {author} {\bibfnamefont {V.}~\bibnamefont {Petrykin}},
  \bibinfo {author} {\bibfnamefont {P.}~\bibnamefont {Krtil}}, \ and\ \bibinfo
  {author} {\bibfnamefont {J.}~\bibnamefont {Rossmeisl}},\ }\href {\doibase
  10.1039/C4CP00571F} {\bibfield  {journal} {\bibinfo  {journal} {Phys. Chem.
  Chem. Phys.}\ }\textbf {\bibinfo {volume} {16}},\ \bibinfo {pages} {13682}
  (\bibinfo {year} {2014})}\BibitemShut {NoStop}%
\bibitem [{\citenamefont {Ohzuku}\ \emph {et~al.}(2016)\citenamefont {Ohzuku},
  \citenamefont {Ikeno}, \citenamefont {Yamada},\ and\ \citenamefont
  {Yagi}}]{Ohzuku2016}%
  \BibitemOpen
  \bibfield  {author} {\bibinfo {author} {\bibfnamefont {H.}~\bibnamefont
  {Ohzuku}}, \bibinfo {author} {\bibfnamefont {H.}~\bibnamefont {Ikeno}},
  \bibinfo {author} {\bibfnamefont {I.}~\bibnamefont {Yamada}}, \ and\ \bibinfo
  {author} {\bibfnamefont {S.}~\bibnamefont {Yagi}},\ }\href {\doibase
  10.1063/1.4961353} {\ \textbf {\bibinfo {volume} {040005}},\ \bibinfo {pages}
  {040005} (\bibinfo {year} {2016})}\BibitemShut {NoStop}%
\bibitem [{\citenamefont {Lodi}\ \emph {et~al.}(1978)\citenamefont {Lodi},
  \citenamefont {Sivieri}, \citenamefont {{De Battisti}},\ and\ \citenamefont
  {Trasatti}}]{Lodi1978}%
  \BibitemOpen
  \bibfield  {author} {\bibinfo {author} {\bibfnamefont {G.}~\bibnamefont
  {Lodi}}, \bibinfo {author} {\bibfnamefont {E.}~\bibnamefont {Sivieri}},
  \bibinfo {author} {\bibfnamefont {A.}~\bibnamefont {{De Battisti}}}, \ and\
  \bibinfo {author} {\bibfnamefont {S.}~\bibnamefont {Trasatti}},\ }\href
  {\doibase 10.1007/BF00617671} {\bibfield  {journal} {\bibinfo  {journal}
  {Journal of Applied Electrochemistry}\ }\textbf {\bibinfo {volume} {8}},\
  \bibinfo {pages} {135} (\bibinfo {year} {1978})}\BibitemShut {NoStop}%
\end{thebibliography}%

\end{document}